\documentclass{article}

\usepackage{arxiv}

\usepackage[utf8]{inputenc} 
\usepackage[T1]{fontenc}    
\usepackage{hyperref}       
\usepackage{url}            
\usepackage{booktabs}       
\usepackage{amsfonts}       
\usepackage{nicefrac}       
\usepackage{microtype}      
\usepackage{lipsum}		
\usepackage{graphicx}
\usepackage[numbers]{natbib}
\usepackage{doi}

\title{A Mechanistic Transform Model for Synthesizing Eye Movement Data with Improved Realism}

\author{ \href{https://orcid.org/0000-0002-4592-8609}{\includegraphics[scale=0.06]{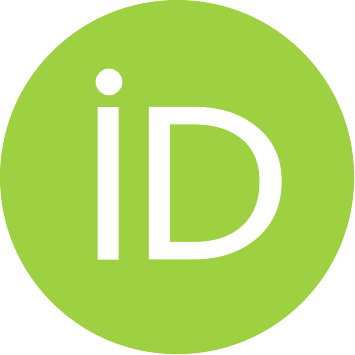}\hspace{1mm}Henry Griffith} \\
	Department of Computer Science\\
	Texas State University\\
	San Marcos, TX 78666 \\
	\texttt{hkgriffith1@gmail.com} \\
	\And
	\href{https://orcid.org/0000-0002-7656-2662}{\includegraphics[scale=0.06]{orcid.pdf}\hspace{1mm}Samantha Aziz} \\
	Department of Computer Science\\
	Texas State University\\
	San Marcos, TX 78666 \\
	\texttt{sda69@txstate.edu} \\
	\And
	\href{https://orcid.org/0000-0002-8088-9270}{\includegraphics[scale=0.06]{orcid.pdf}\hspace{1mm}Dillon J Lohr} \\
	Department of Computer Science\\
	Texas State University\\
	San Marcos, TX 78666 \\
	\texttt{djl70@txstate.edu} \\
    \And	\href{https://orcid.org/0000-0001-7890-8842}{\includegraphics[scale=0.06]{orcid.pdf}\hspace{1mm}Oleg Komogortsev} \\
	Department of Computer Science\\
	Texas State University\\
	San Marcos, TX 78666 \\
	\texttt{ok@txstate.edu} \\
}

\renewcommand{\shorttitle}{A Mechanistic Transform Model for Synthesizing Eye Movement Data}

\begin{document}
\maketitle

\begin{abstract}
 This manuscript demonstrates an improved model-based approach for synthetic degradation of previously captured eye movement signals. Signals recorded on a high-quality eye tracking sensor are transformed such that their resulting eye tracking signal quality is similar to recordings captured on a low-quality target device. 
 The proposed model improves the realism of the degraded signals versus prior approaches by introducing a mechanism for degrading spatial accuracy and temporal precision. 
 Moreover, a percentile-matching technique is demonstrated for mimicking the relative distributional structure of the signal quality characteristics of the target data set.
 The model is demonstrated to improve realism on a per-feature and per-recording basis using data from an EyeLink 1000 eye tracker and an SMI eye tracker embedded within a virtual reality platform. 
 The model improves the median classification accuracy performance metric by 35.7\% versus the benchmark model towards the ideal metric of 50\%. 
This paper also expands the literature by providing an application-agnostic realism assessment workflow for synthetically generated eye movement signals.
\end{abstract}


\section{Introduction}
Eye tracking (ET) technology has been deployed in laboratory settings for various value-added applications for decades, including health assessment~\cite{snegireva2018eye}, education~\cite{ashraf2018eye}, and marketing~\cite{wedel2017review}. More recently, ET sensors are being utilized within consumer electronic devices as an input modality~\cite{bazrafkan2015eye}, including head-mounted devices for mixed reality applications~\cite{gardony2020eye}. As the form-factor of the target integration platform changes, the corresponding eye tracking signal quality (e.g.: spatial accuracy and precision, temporal precision, etc.~\cite{holmqvist2012eye}) provided by the ET sensor also typically varies~\cite{imaoka2020assessing,raynowska2018validity}, thereby limiting the generalization of signal processing workflows across hardware platforms. 

To help address the design challenges associated with variability in signal quality across ETs, various model-based techniques for transforming ET signals have been previously proposed (e.g.~\cite{zemblys2018using,otero2014unsupervised}, etc.). These algorithms allow for data recorded using a particular ET device (hereby denoted as the source) to appear as if they were instead recorded on a different device (hereby denoted as the target). These transformations are valuable for synthetically augmenting the scale and diversity of available training data for machine learning applications (e.g. ~\cite{zemblys2018using}). In addition, they allow system designers to estimate performance variability for various target applications as a function of ET sensor quality (e.g. ~\cite{lohr2019evaluating}). Such transformations may either explicitly utilize the performance characteristics of the target hardware as parameters (e.g.: the simplistic downsampling of data to a target rate, etc.), or develop estimates of the underlying hardware style by modeling a representative target data set.

The synthetic transformation of eye tracking signals is complicated by several factors. Namely, the metrics and corresponding definitions used to assess device quality vary considerably across the research community (e.g.~\cite{lohr2019evaluating}, etc.). In addition, variability in eye tracking signals from their expected value may be associated with both noise from the recording instrument, along with noise from the human visual system (HVS) itself~\cite{st1973signal}. To address this latter concern, recent models have suggested injecting noise with a uniform power spectral density, which has been shown to resemble the instrument noise component through the assessment of artificial eye recordings~\cite{coey2012structure}. While this strategy improves the realism of the relative instrument-to-HVS noise levels under the stated assumptions, it does not support the modulation of both the spatial accuracy and temporal precision of the source data. Moreover, it lacks the ability to replicate the distributional structure of the desired performance metrics across a target data set. 

\begin{figure}
\centering
\includegraphics[width=0.7\linewidth]{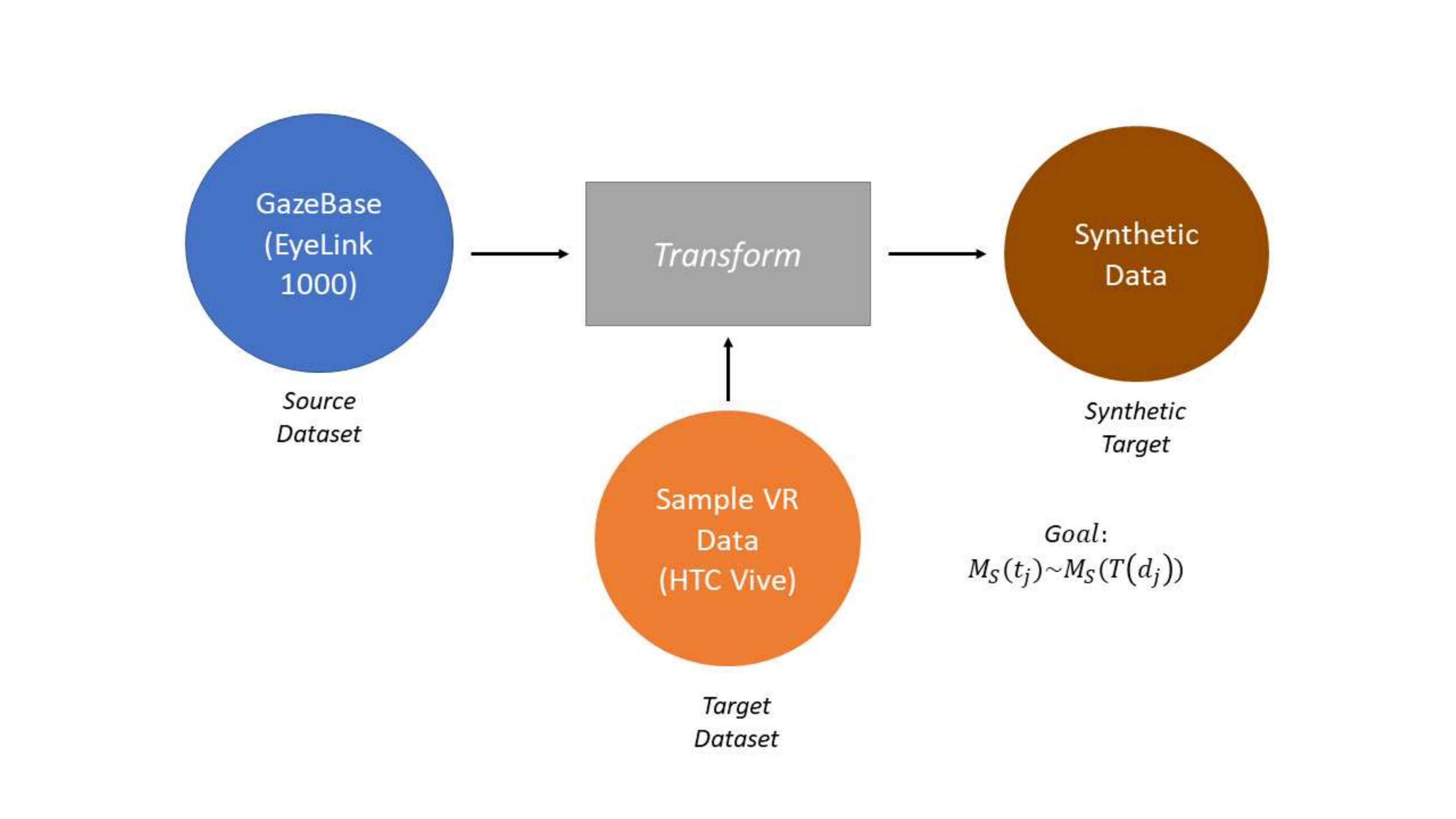}
\caption{System diagram of transformation workflow. GazeBase data serves as the source data set, which is transformed using a model-based approach to yield synthetic data. Transform parameters are informed by the target data set to yield similarity in eye tracking signal quality characteristics across the synthetic and target data sets.}
\label{fig:SystemDiagram*}
\end{figure}

The research described herein demonstrates a series of improvements for a previously proposed additive white noise degradation model ~\cite{zemblys2018using} which enhances the realism of the synthetically generated signals. Namely, modifications are introduced to support the marginal degradation of both the spatial accuracy and temporal precision of the source data set. Moreover, a percentile-matching technique is proposed in order to mimic the underlying distributional structure of the target data set quality metrics across recordings. 
The proposed technique is validated using subsets of GazeBase~\cite{griffith_gazebase_figshare2020} as the source data set and GazeBaseVR~\cite{Lohr2022} as the target data set.
The realism of the transformation is assessed by comparing quality metrics computed on the synthetically-generated data to those computed on the target data on both a per-feature and per-recording basis. The corresponding processing workflow employed herein is summarized in Figure~\ref{fig:SystemDiagram*}. 

The primary contributions of this work include - 1) the introduction and demonstration of a modified mechanistic transform model which offers enhanced parameterization and realism for the model-based synthetic degradation of eye tracking signals, and 2) the proposal and demonstration of an application-agnostic assessment workflow suitable for assessing the realism of synthetically generated signals on a per eye tracking signal quality metric and per-recording level using a 1-nearest neighbor (NN) classifier. While the efficacy of this assessment technique has been previously demonstrated for images generated using generative adversarial networks (e.g.~\cite{xu2018empirical}), we are unaware of its application for the assessment of synthetically-generated eye movement signals. 

\section{Methods}
\subsection{Data}
A subset of the publicly-available GazeBase repository~\cite{griffith_gazebase_figshare2020} was utilized as the source data within this project. While a full description of the collection protocol for GazeBase is provided in~\cite{griffith2020gazebase}, an abbreviated description is provided within this manuscript for completeness. GazeBase is a multi-stimuli, large-scale, longitudinal data set of monocular (left) eye movement recordings captured over a 37-month period from 322 college-aged subjects. Each collection period (denoted as a round) consists of a pair of contiguous recording sessions, during which subjects completed a battery of common tasks across both sessions. Tasks included fixation, random saccade, video viewing, reading, and interactive game play. A subset of subjects from each prior round completed subsequent rounds of recording, with a total of nine rounds of recording performed.

Eye movements were captured in GazeBase using the EyeLink 1000 eye tracker at 1,000 Hz~\cite{sr2010eyelink}. For the current analysis, only Round 1 random saccade data from participants with subject identifiers less than 100 (99 total participants, 198 files across both sessions) were utilized. The random saccade task was chosen as it represented the most complex guided-viewing task (e.g.: task with specified stimulus locations) available within the collection. Guided-viewing tasks are necessary due to limitations in the definitions of the employed eye tracking signal quality metrics of the target data set as described in~\cite{lohr2019evaluating}.

The target data employed herein was captured using a subset of the publicly-available GazeBaseVR repository~\cite{Lohr2022}.
The experimental battery for this collection was chosen to align with the GazeBase collection procedure as much as possible (e.g.: utilization of similar tasking, contiguous recording sessions within individual rounds, etc.).
Similar to GazeBase, GazeBaseVR is a multi-stimuli, longitudinal data set of binocular eye movement recordings captured over a 26-month period from 407 college-ages subjects.
Each round comprises a a pair of contiguous recording sessions where subjects completed a battery of five tasks.
Tasks for GazeBaseVR included vergence, smooth pursuit, video viewing, reading, and random saccade. 
Subjects completed up to a total of 3 rounds; while subjects present in Rounds 2 and 3 are present in Round 1, a subject in Round 3 may not be present in Round 2 and vice versa. 

Eye movements in GazeBaseVR were captured using a modified HTC Vive with an embedded SMI eye-tracking device at 250~Hz~\cite{Lohr2022}. 
Recordings from 276 subjects captured while completing the Round 1 random saccade task were utilized herein. It should be explicitly noted that both the subject pool and target (stimulus) trajectories of the random saccade task in GazeBaseVR are disjoint from those utilized in GazeBase. Namely, targets remained stationary during the random saccade task for a fixed period of time (1000~ms) for GazeBase and for a random period of time (uniformly distributed from 1000--1500~ms) for GazeBaseVR. In addition, the target pattern was varied across the two experiments, and no common subjects were utilized across the two collections. As the technique described herein compares signals in terms of data quality metrics which are computed upon fixations, the aforementioned differences in stimuli trajectories should not significantly influence the resulting analysis.

Although GazeBaseVR captures binocular gaze data, only data from the left eye was utilized within this analysis. Monocular data is reported by the modified HTC Vive at a nominal sampling rate of 250 Hz, with some variation around the ideal intersampling interval~(ISI) of 4~ms observed in the resulting data. This variability is in contrast to the data reported by the EyeLink 1000 device, which has a constant ISI of 1~ms. Additional information regarding the achievable quality levels of the HTC Vive platform may be found in~\cite{lohr2019evaluating} and \cite{Lohr2022}. 

\subsection{Eye Tracking Signal Quality Metrics}
Both per-channel and combined spatial accuracy and precision, along with temporal precision, were used to characterize eye tracking signal quality within this workflow at various stages (i.e.: for both informing the transformation model and for assessing its efficacy). These metrics were chosen as they represented the best inherent summary characterization of the eye tracking sensor. Each metric was computed using the definitions described in~\cite{lohr2019evaluating}. Namely, spatial accuracy is defined as the systematic bias of the gaze measurements produced by an eye tracker about the ground truth value, while precision describes the corresponding dispersion of these measurements about their central tendency. Temporal precision describes the variation in the ISI about the nominal value of the device. While full details of the eye tracking signal quality computations are omitted for brevity, the general procedure is outlined below. This content focuses on the process for estimating the temporal domain of each fixation within the random saccade task absent of the use of algorithmic classification.  

\begin{figure}[ht]
\centering
\includegraphics[width=0.5\linewidth]{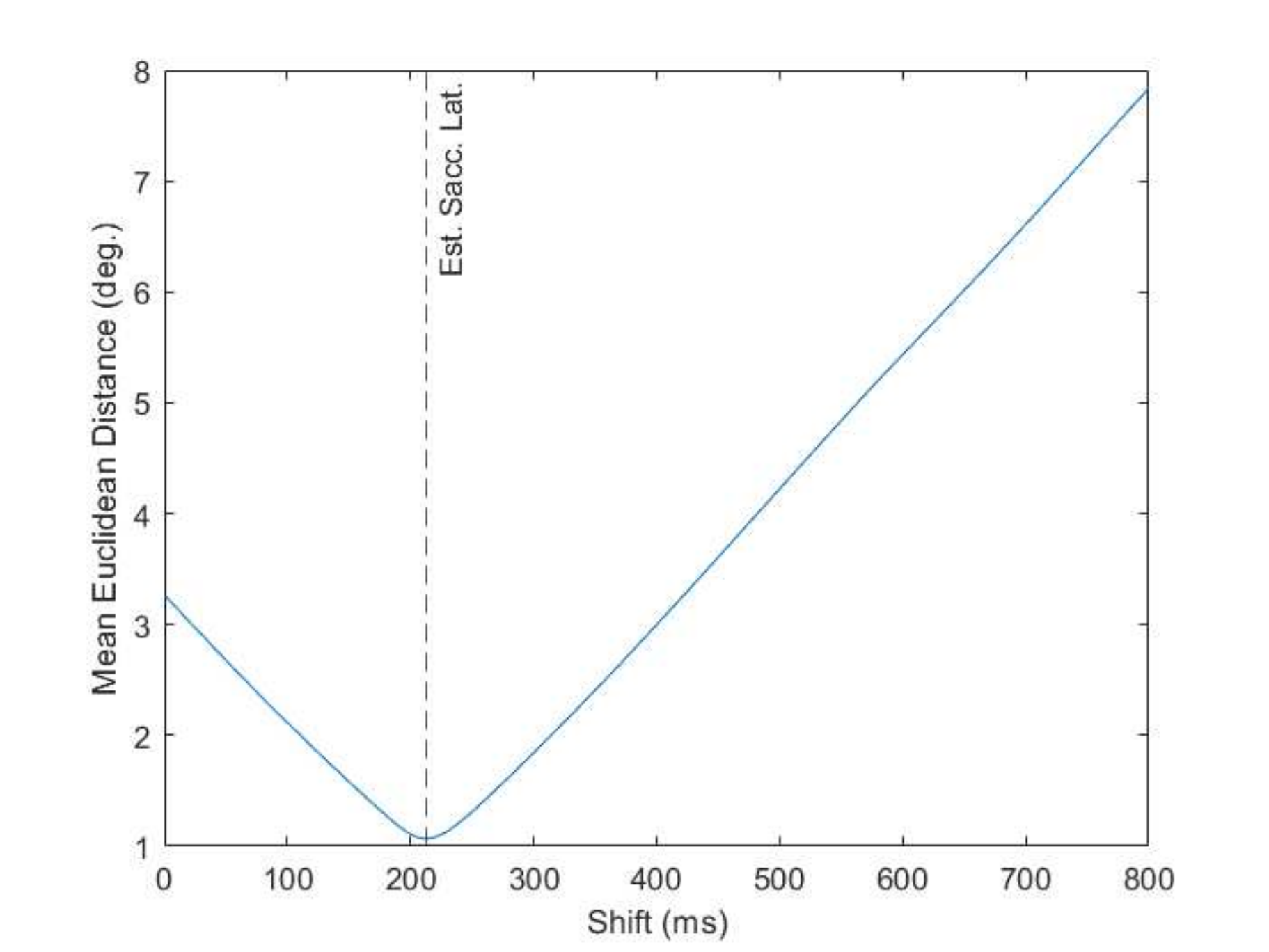}
\caption{Example of variation in Euclidean distance between stimulus and shifted target signal across range of shift values considered for latency removal.}
\label{fig:DistVsShift}
\end{figure}

The computation of eye tracking signal quality metrics was initiated by estimating the average saccade latency on a per-file basis in order to parse the fixation intervals on each stimulus. Saccade latency accounts for the delay between stimulus transitions and the corresponding saccade initiation, which is typically on the order of 200 ms~\cite{leigh2015neurology}. Saccade latency was estimated by finding the shifted value of the gaze signal which exhibited minimum separation from the target signal in the Euclidean sense. An example of the variability in Euclidean distance over the range of considered latency values is shown in Figure ~\ref{fig:DistVsShift}.

\begin{figure*}[ht]
\centering
\includegraphics[width=0.9\linewidth]{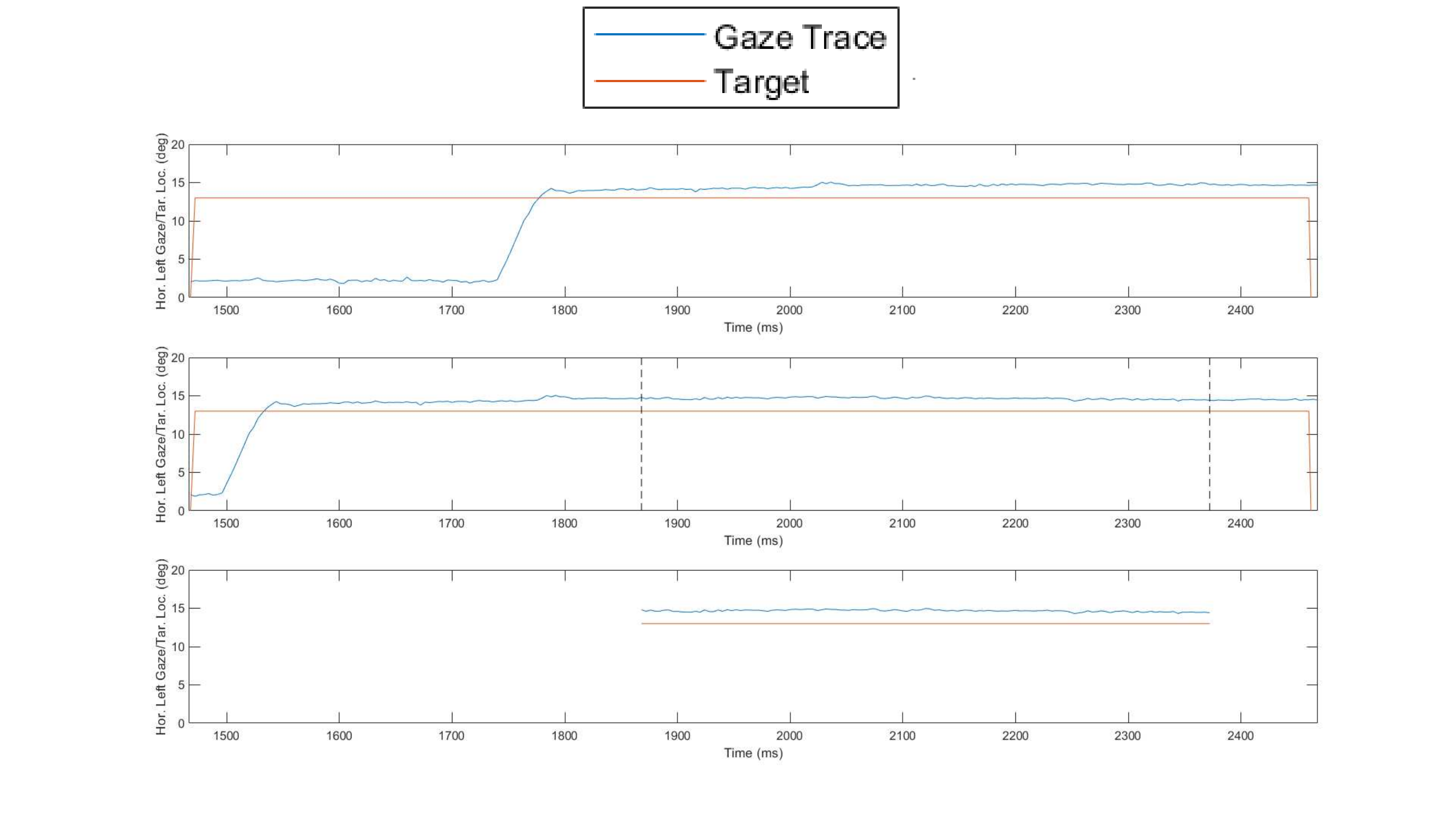}
\caption{Example of latency alignment and partitioning for candidate fixations. Top: Original gaze and target signal. Middle: Target signal with gaze signal shifted by per-file latency estimate. Note that the corresponding alignment is not ideal due to inter-recording variability in saccade latency. Dashed vertical lines represent the start and stop points of the partitioning process, which are 400 and 900 ms after the target transition point, respectively. Bottom: Candidate fixation resulting from the extraction procedure. Note that by discarding the first 400 ms, the fixation in the adjusted gaze signal is removed, thereby yielding a relatively stable fixation signal for computing the data quality metrics.}
\label{fig:Extraction}
\end{figure*}

Once the saccade latency had been estimated on a per-recording basis, the start and stop times for each fixation were estimated based upon adjusting the target transition times according to the estimated latency. Next, the first 400 ms after the estimated fixation start were discarded in order to account for inter-recording variability in saccade latency. Namely, the removal of this portion of the signal helps ensure that only valid fixation intervals are captured for subsequent computation of the data quality metrics. Once this initial offset was discarded, the following 500 ms was selected for subsequent processing. An example of this aggregate extraction procedure is shown in Figure ~\ref{fig:Extraction}. 

Within each candidate fixation, samples with a distance-to-centroid either outside Tukey's fences~\cite{tukey1977exploratory} or greater than 2~degrees of the visual angle (dva) were marked as outliers and discarded from further analysis. Once outliers had been removed, the spatial accuracy of each fixation was computed on a per-channel and combined basis according to equations \ref{eq:acc-x}--\ref{eq:acc-c}.

\begin{equation}
    \theta_h = \frac{1}{n}\sum_{i=1}^{n} {| x^g_i - x^t_i |}
    \label{eq:acc-x}
\end{equation}
\begin{equation}
    \theta_v = \frac{1}{n}\sum_{i=1}^{n} {| y^g_i - y^t_i |}
    \label{eq:acc-y}
\end{equation}
\begin{equation}
    \theta_c = \frac{1}{n}\sum_{i=1}^{n} \sqrt{(x^g_i - x^t_i)^2 + (y^g_i - y^t_i)^2}
    \label{eq:acc-c}
\end{equation}
where $\theta_h $, $\theta_v $, and $\theta_c $ correspond to the horizontal, vertical, and combined spatial accuracy in dva, respectively; $x^g$ and $x^t$ correspond to the gaze and target samples in the horizontal channel, respectively; $y^g$ and $y^t$ are analogous to $x^g$ and $x^t$ for the vertical channel; and $n$ corresponds to the number of samples within the fixation. In a similar fashion, the spatial precision of each fixation was computed on as shown in equations \ref{eq:pre-x}--\ref{eq:pre-c}.

\begin{equation}
    MAD_h =M({| x^g_i - M(x^g) |})
    \label{eq:pre-x}
\end{equation}
\begin{equation}
    MAD_v =M({| y^g_i - M(y^g) |})
    \label{eq:pre-y}
\end{equation}
\begin{equation}
    MAD_c = {\sqrt{M(| x^g_i - M(x^g) |)^2 + M(| y^g_i - M(y^g) |)^2}}
    \label{eq:pre-c}
\end{equation}
where $M() $ denotes the median operator, and $MAD_h $, $MAD_v $, and $MAD_c $ denote the horizontal, vertical, and combined spatial precision, respectively.

Temporal precision was computed by taking the standard deviation of the sample-over-sample difference in timestamps on a per-file basis. 

\subsection{Benchmark Transformation Model} \label{subsec:details}
The mechanistic data degradation model described in~\cite{zemblys2018using} was implemented herein for initial benchmarking purposes. While it was attempted to replicate the implementation as closely as possible to the procedure described in the original manuscript, some modifications were made to enhance  performance for the current application as described in the remainder of this section. This modified model is hereby denoted as the benchmark model. 

The benchmark model implements independent mechanisms to reduce bandwidth and degrade the spatial precision of the source data set. Bandwidth is reduced through a customized downsampling procedure intended to avoid aliasing in the downsampled signal. In the initial implementation of the benchmark model, the source signal was filtered using a Butterworth filter with a cutoff rate set at 0.8 times the Nyquist frequency associated with the target sampling rate. For the current application, this filter was modified to a zero-phase architecture to minimize the delay between the filtered and original signal. After filtering, the signal is resampled at the target sampling rate using first-order spline interpolation. 

Spatial precision is degraded in the benchmark model using an additive Gaussian noise stationary process as shown in equations \ref{eq:add-noise-x}--\ref{eq:noise-def}. As noted in equation \ref{eq:sigma-def}, the variance of the additive noise is tuned to increase towards the screen edges using a Gaussian weighting function. 

\begin{equation}
    x_t =x_s + N_x
    \label{eq:add-noise-x}
\end{equation}
\begin{equation}
    y_t =y_s + N_y
    \label{eq:add-noise-y}
\end{equation}
\begin{equation}
    N_x, N_y \sim \mathcal{N}(0,\,\sigma^2(x_s,\,y_s))
    \label{eq:noise-def}
\end{equation}
\begin{equation}
    \sigma^2(x_s,\,y_s)=\alpha(x_s,\,y_s) \times \sigma_0^2
    \label{eq:sigma-def}
\end{equation}
\begin{equation}
    \alpha(x_s,\,y_s)=\exp{\left(\frac{-(r_s-r_{max})^2}{2\sigma_s^2}\right)}
    \label{eq:exp}
\end{equation}
where $x_t$ and $x_s$ correspond to the synthetic target and real source samples in the horizontal channel, respectively; $y_t$ and $y_s$ are analogous to $x_t$ and $x_s$ for the vertical channel; $\sigma_0$ denotes the maximum value of the additive noise variance; $\sigma_s$ denotes the dispersion parameter in the Gaussian weighting function; $r_{max}$ denotes the maximum radial dimension of the screen in dva; and $r_s=\sqrt{x^2_s+y^2_s}$. As preliminary analysis of the target data utilized within this work demonstrated no variability in precision as a function of eccentricity, the scaling function $\alpha$ was removed from the original algorithm, thereby yielding the modified distributional structure described in equation \ref{eq:modNoise}.

\begin{equation}
    N_x, N_y \sim \mathcal{N}(0,\,\sigma_0^2)
    \label{eq:modNoise}
\end{equation}

\subsection{Benchmark Model Tuning}\label{sec:Benchmark}
\begin{figure}[ht]
\centering
\includegraphics[width=0.5\linewidth]{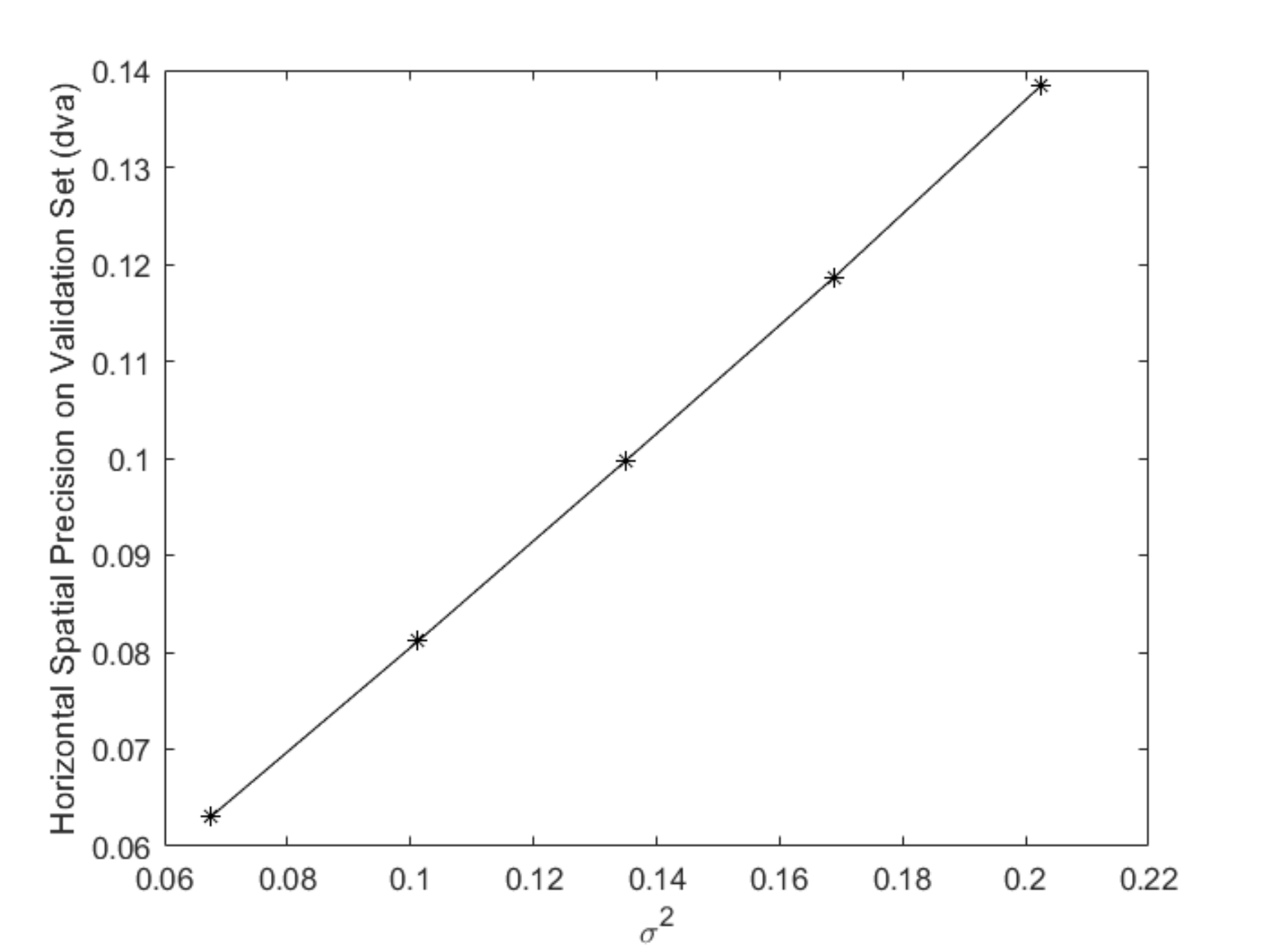}
\caption{Variability in horizontal spatial precision for transformed source data versus $\sigma^2_0$.}
\label{fig:ValTuning}
\end{figure}

Tuning the benchmark model in its native implementation requires specification of the nominal sampling rate of the target eye tracker (250 Hz), along with determination of the additive noise variance parameter $\sigma^2_0$. This latter value was empirically determined for the current analysis by evaluating the horizontal spatial precision computed on the source data set of transformed data for varying values of $\sigma^2_0$ as depicted in Figure ~\ref{fig:ValTuning}. As shown, a value of $\sigma^2_0=0.13$ yielded a horizontal spatial precision value which is approximately equal to the observed value in the target data set (e.g. $MAD_h = 0.097 $).

\subsection{Development of Modified Degradation Model}
\begin{figure}[ht]
\centering
\includegraphics[width=0.8\linewidth]{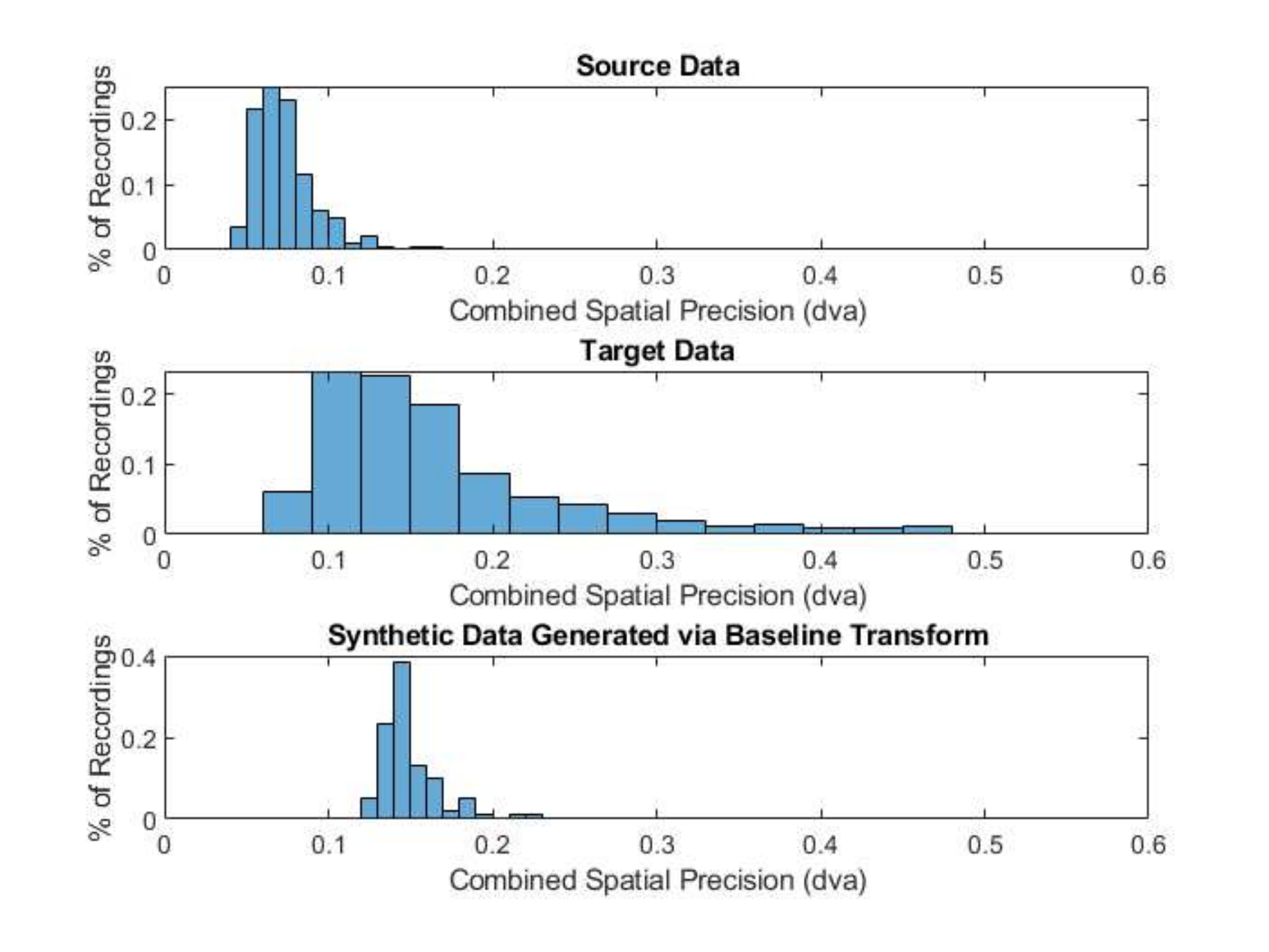}
\caption{Distribution of combined spatial precision across files for the target (top) and synthetically degraded  sample (bottom). As noted, while the benchmark model matches the target spatial precision in central tendency, the shape of the distribution is not maintained.}
\label{fig:BadPrecision}
\end{figure}

The baseline model is characterized by several notable limitations. Namely, as the additive noise parameter utilized is applied constantly across each file in the source data set, the corresponding distribution of the spatial precision of generated target data only matches that of the source data in the central tendency set (e.g.: the relative dispersion across files is distorted). This phenomenon is demonstrated in Figure~\ref{fig:BadPrecision} for the case of combined spatial precision, where the resulting synthetic data retains the distribution of the source data with its central tendency shifted to match the target data. In addition to this limitation, as the baseline algorithm does not marginally degrade either the spatial accuracy or temporal precision of the source data set, the resulting synthetic data is not capable of exhibiting resemblance to the target with respect to these measures.

\begin{figure}[ht]
\centering
\includegraphics[width=\linewidth]{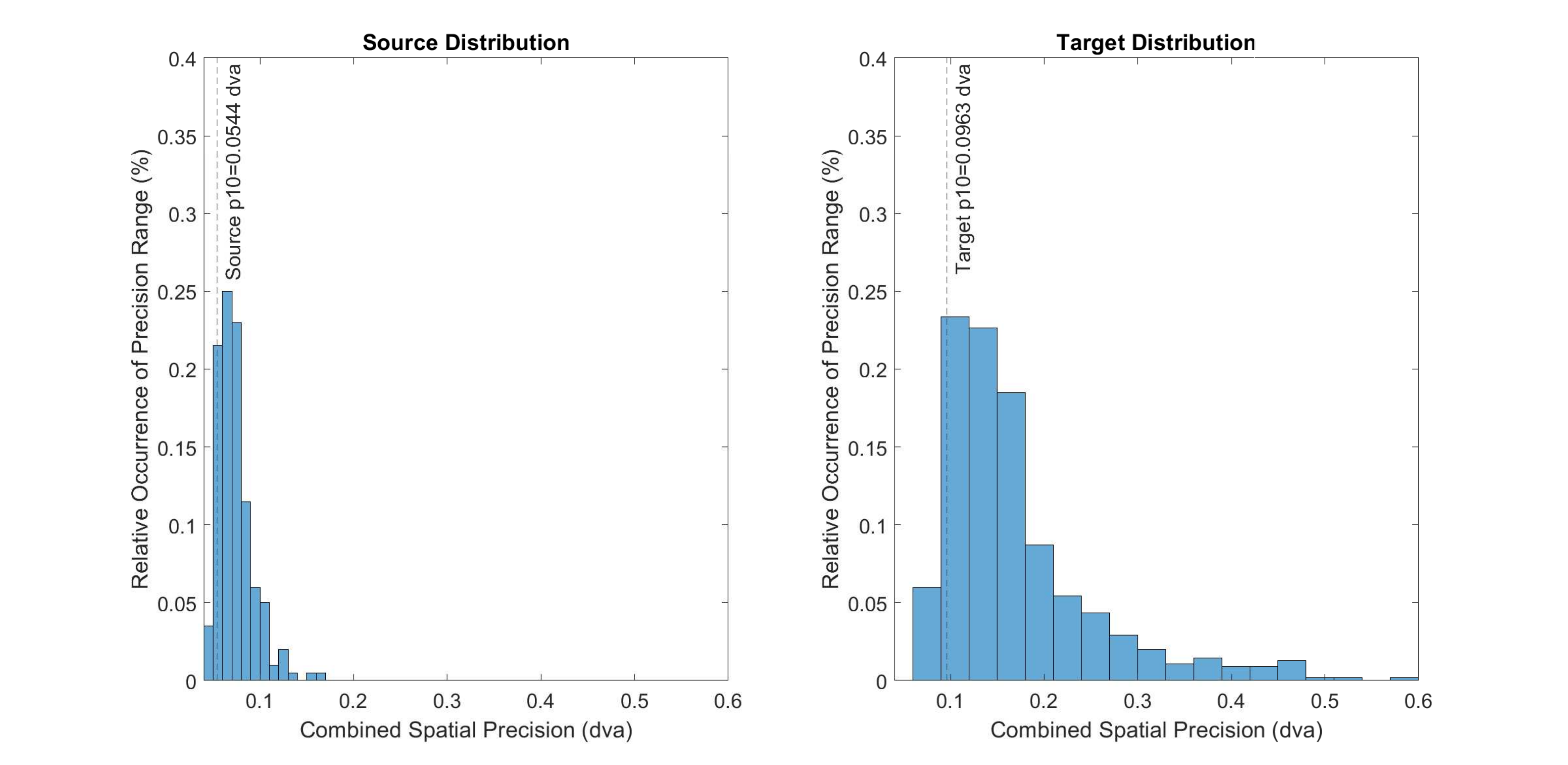}
\caption{Proposed technique for maintaining the underlying distributional structure of a eye tracking signal quality metric across recordings (shown for the case of combined spatial precision). For each file in the source set, combined spatial precision is initially computed and its relative percentile rank within the source set is determined. Based upon matching this percentile value in the target data set, the target spatial precision value is determined, with the requisite tuning parameter (i.e.: $\sigma^2_0$) to achieve the marginal precision degradation subsequently estimated.}
\label{fig:PercentileMatching}
\end{figure}

To address these limitations of the baseline model, various modifications were implemented. To retain the distributional structure of the spatial precision of the target data set, a percentile-matching technique was implemented as shown in Figure~\ref{fig:PercentileMatching}. In this technique, the percentile ranking of the combined spatial precision within the source distribution for each file is initially computed. The spatial precision of the matching percentile within the target distribution was then determined in order to compute the requisite marginal precision degradation. Based upon this value, the requisite additive noise parameter was determined using the linear regression model discussed in Section ~\ref{sec:Benchmark} in the marginal sense. 

A similar percentile-matching approach was also utilized to marginally degrade the spatial accuracy of the source signals to resemble the corresponding distributional within the target data set. Namely, the marginal accuracy degradation for a given source file was initially determined on a per-channel basis. Then, for each fixation within the source file, a specific per-channel degradation value was drawn from a normal distribution centered around the value computed in the previous step. The standard deviation of this distribution was chosen such that 99.7\% of the values were within 20\% of the requisite accuracy degradation value. The sampled value was subsequently added to the gaze signal during the fixation after weighting by a uniformly distributed random sign (e.g.: $+/-$) value. This randomness (e.g.: both variability from the requisite marginal degradation value and in offset direction) was used as a na\"{i}ve initial approach to improve realism of the resulting synthetic signals. 

\begin{figure}[ht]
\centering
\includegraphics[width=0.55\linewidth]{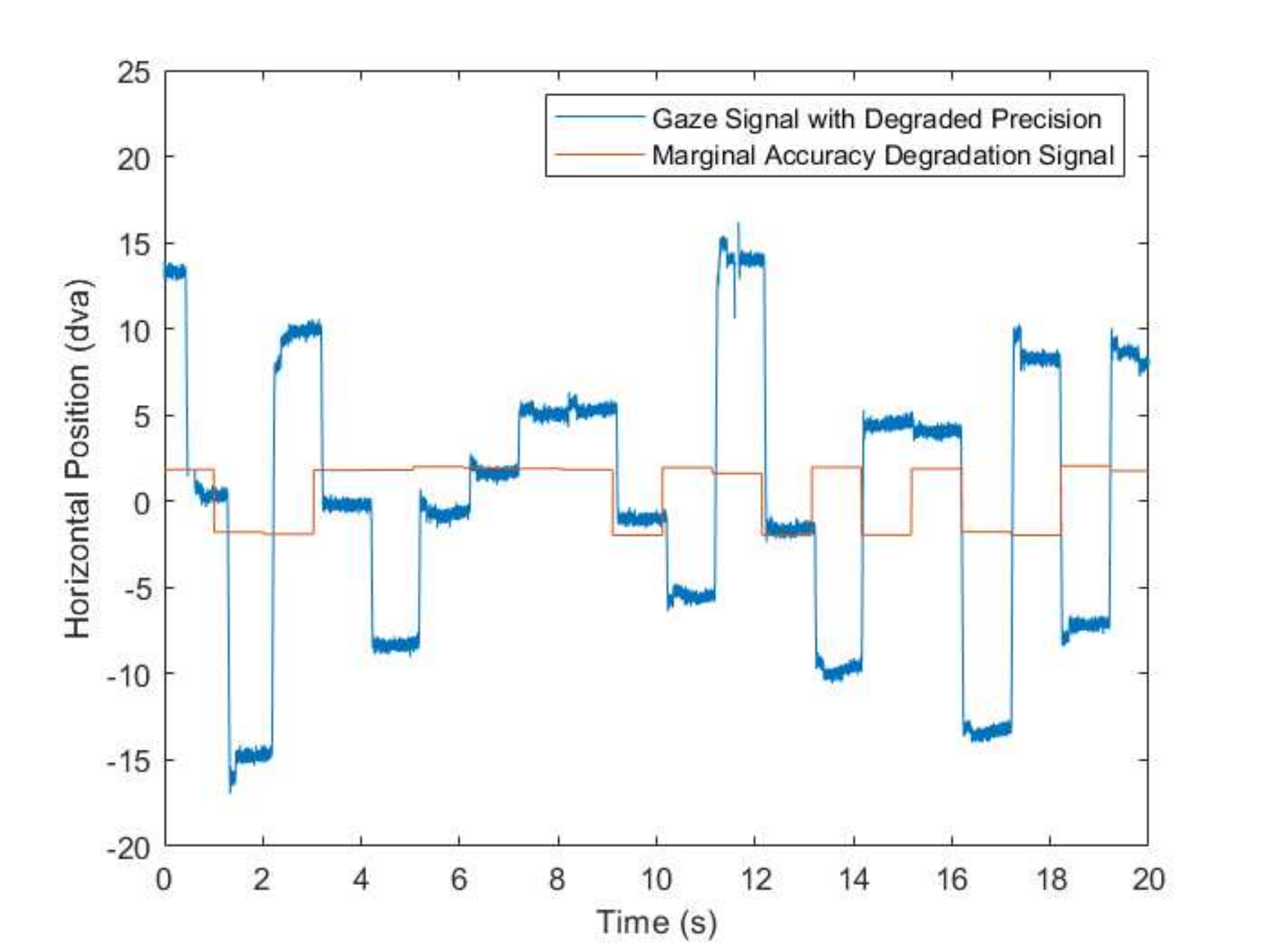}
\caption{Example of the marginal accuracy degradation signal used to degrade the spatial accuracy of a source signal to its percentile match in the target distribution. This procedure (shown only for the horizontal channel within the figure) was replicated on a per-channel basis.}
\label{fig:MarginalAccuracyDeg}
\end{figure}

To determine fixation boundaries for adding the marginal accuracy degradation signal, a shift-based alignment approach similar to that described above and in~\cite{lohr2019evaluating} was employed (e.g..: the marginal accuracy degradation signal was added to the latency-adjusted gaze signal which was defined as the shifted gaze signal which minimized the Euclidean distance computed between the target and gaze value). An example of the marginal degradation signal utilized within a given recording is depicted in Figure~\ref{fig:MarginalAccuracyDeg}.

To replicate the degradation in temporal precision observed in the target data set, the domain values for the re-sampling procedure in the original model were perturbed by adding a zero-mean random Gaussian noise term to each nominal time stamp. The standard deviation of this distribution was set equal to the empirical value observed variation in the ISIs from the nominal sampling period (i.e.: 4 ms) of the target set. 

The various changes implemented within the modified Zemblys et al.~\cite{zemblys2018using} algorithm are summarized below. 

\begin{itemize}
  \item A percentile-matching technique was utilized to attempt to match the distributional structure of the spatial precision observed in the target data set (e.g. Figure ~\ref{fig:PrecFig}).
  \item A technique for degrading the spatial accuracy was implemented.The percentile matching technique described above was then utilized to mimic the distributional structure of this quality parameter in the target data set;
  \item A technique for degrading the temporal precision of the algorithm was implemented. 
\end{itemize}

\begin{figure}[ht]
\centering
\includegraphics[width=0.55\linewidth]{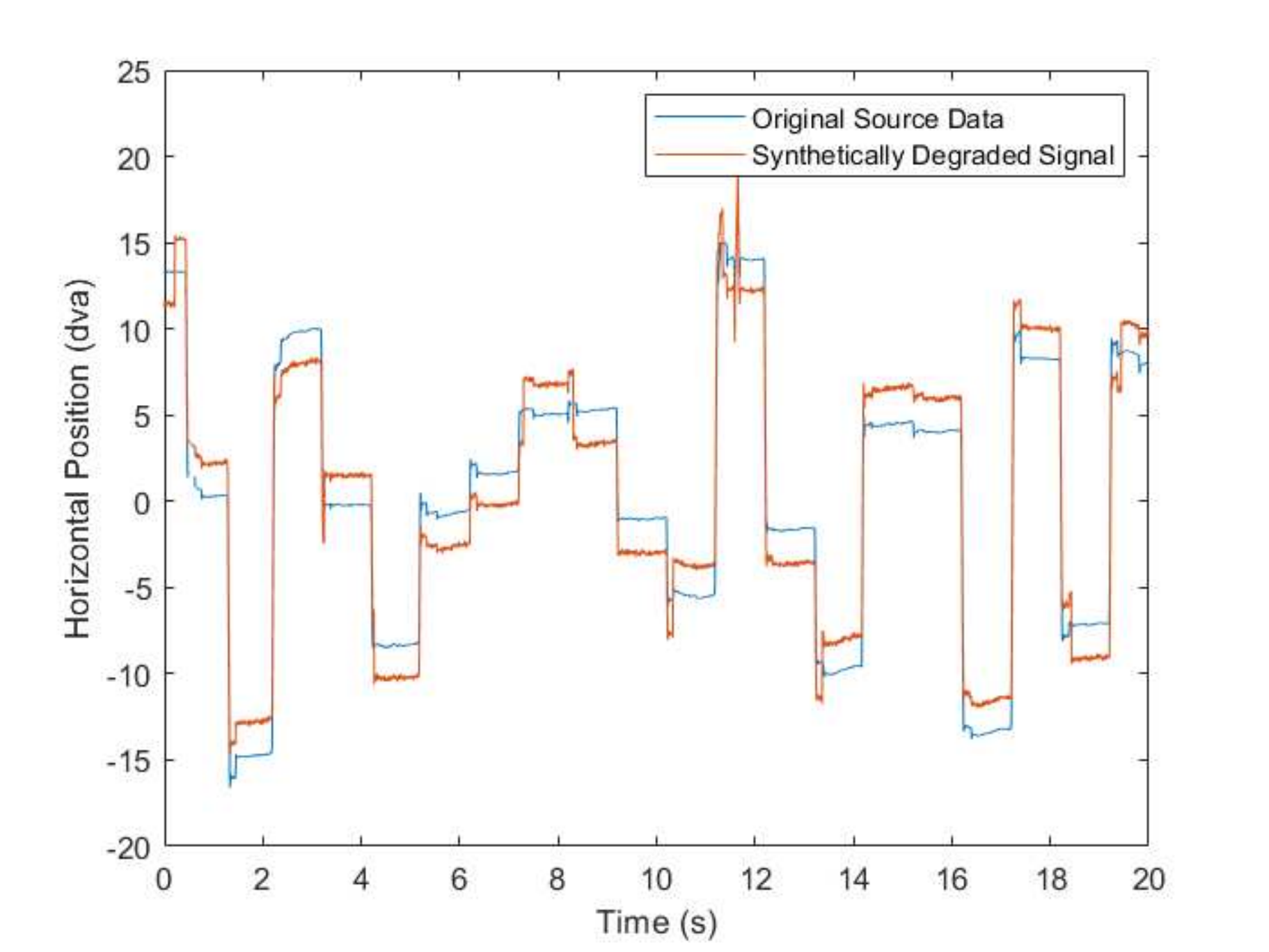}
\caption{Example of a source and synthetically degraded target signal produced by the proposed algorithm.}
\label{fig:ExSigs}
\end{figure}

An example of the result of applying this aggregate workflow is shown in Figure~\ref{fig:ExSigs}.

\section{Results}
\subsection{Comparison of Individual Eye Tracking Signal Quality Metrics}
\begin{figure}[ht]
\centering
\includegraphics[width=\linewidth]{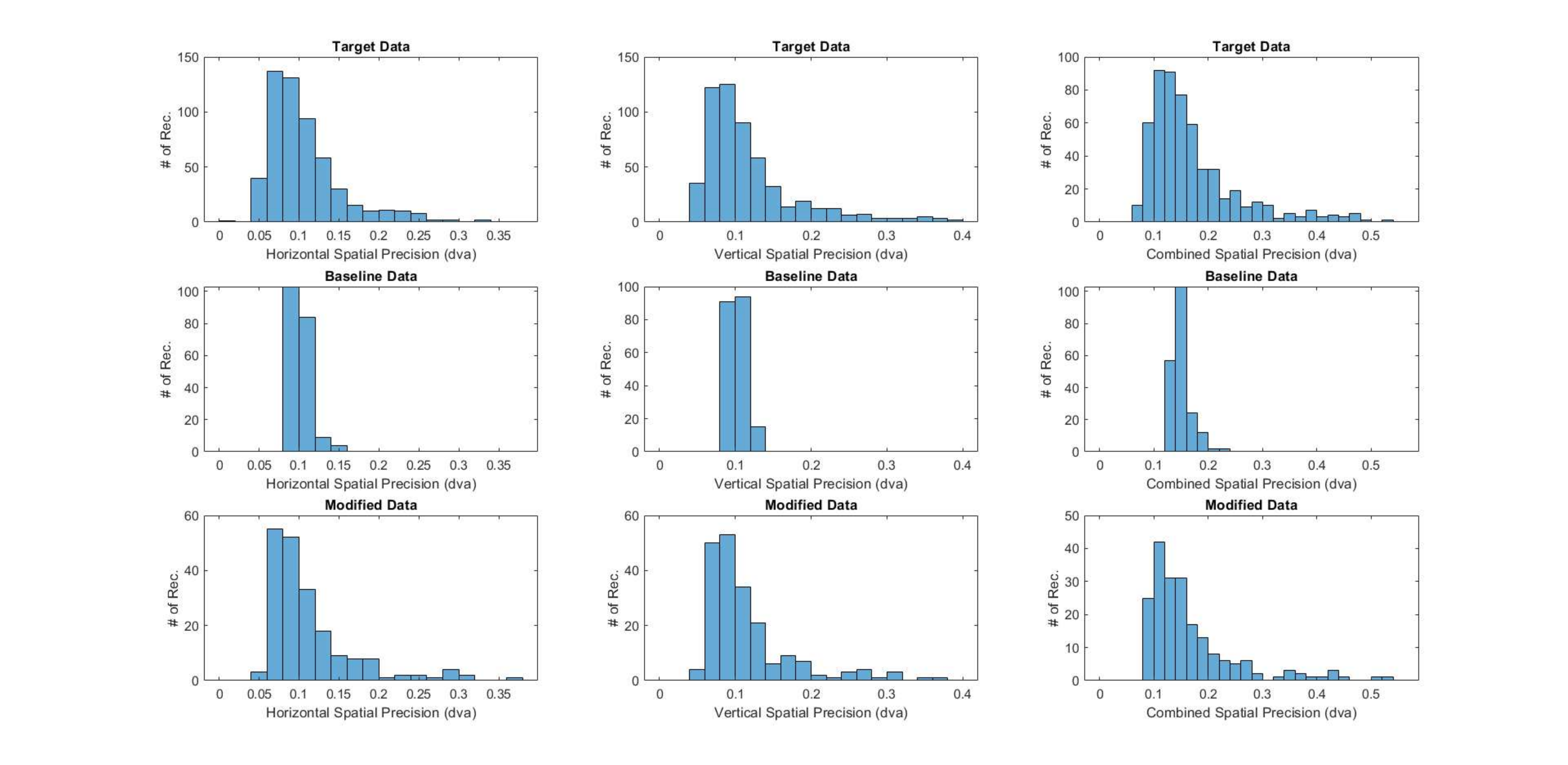}
\caption{Variation in spatial precision eye tracking signal quality metrics across the target and synthetic data sets produced by the baseline and modified model, respectively.}
\label{fig:PrecFig}
\end{figure}

\begin{figure}[ht]
\centering
\includegraphics[width=\linewidth]{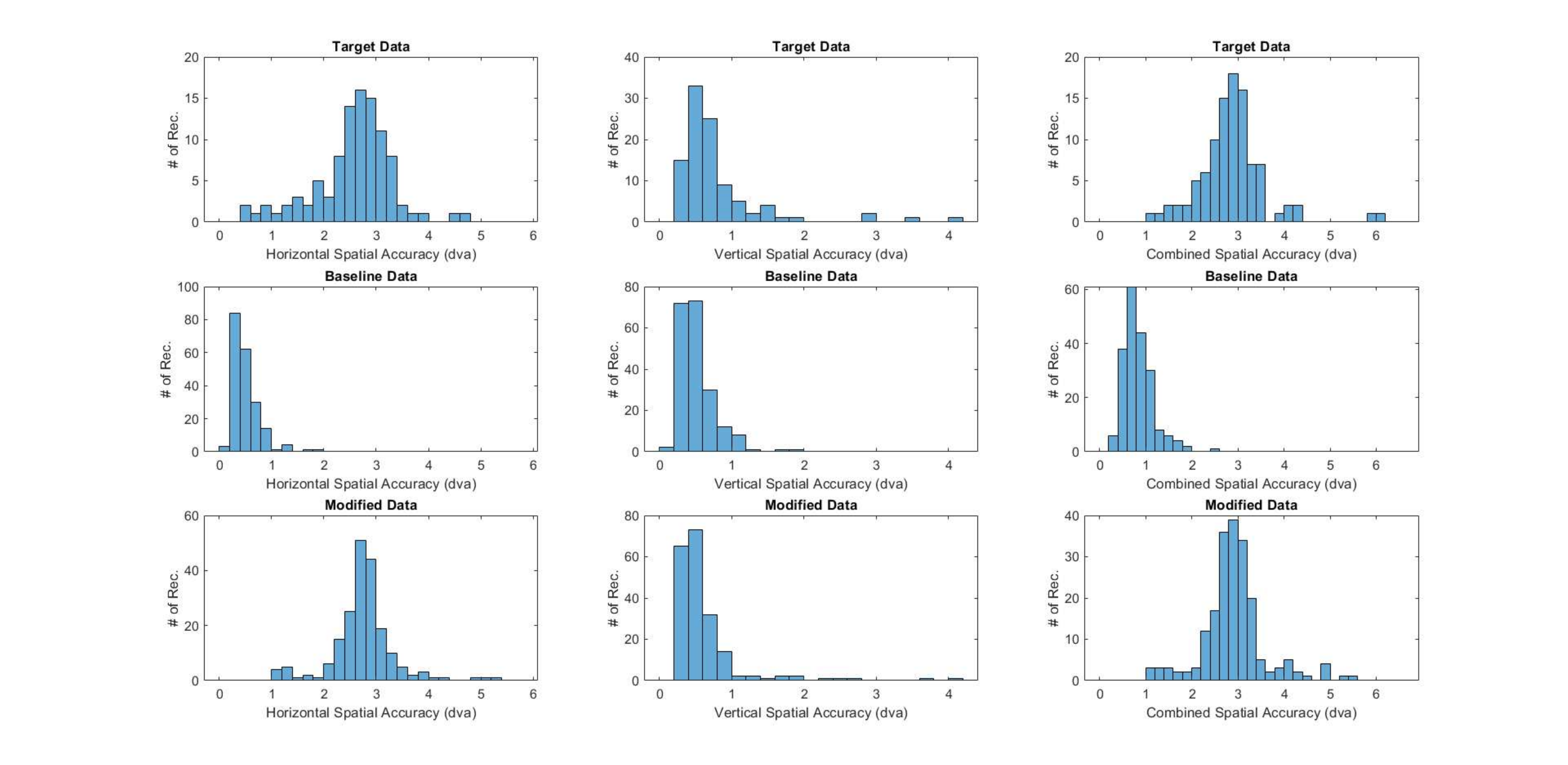}
\caption{Variation in spatial accuracy eye tracking signal quality metrics across the target and synthetic data sets produced by the baseline and modified model, respectively.}
\label{fig:AccFig}
\end{figure}

The eye tracking signal quality of data generated using the proposed and benchmark model was compared to the real target data set in order to assess the efficacy of the two approaches. A visualization of precision-related metrics across the two synthetic data sets and target data set is shown in Figure ~\ref{fig:PrecFig}. A similar presentation of accuracy-related metrics is shown in Figure ~\ref{fig:AccFig}. As shown, the distributions in the synthetic data produced by the modified workflow exhibit significantly improved resemblance to the target data when compared to the baseline model. It should be noted that temporal precision metrics are not visualized since the baseline model has no mechanism for degrading this parameter (i.e.: source data transformed using the baseline model would exhibit no variability in temporal precision, making visual comparison with the target data set trivial). 

\subsection{Per-Recording Realism Assessment}
The realism of the synthetically generated signals was also assessed on an aggregate basis using a 1-NN two-sample test as described in~\cite{xu2018empirical}. Namely, a 1-NN classifier was employed on eye tracking signal quality feature vectors of the target and synthetic data sets using a leave-one-sample-out strategy. For an ideal transform scenario, the corresponding accuracies would be 50\%, corresponding to the inability of the classifier to distinguish between synthetic and real data at a performance level better than chance. 

As this method requires that the size of the two sample sets be identical, a subset of the target data was randomly sampled for comparison. To elucidate the effect of this sampling, the resulting analysis performed herein was repeated five times for random samples. An identical analysis was also performed on the baseline model. Classification results are summarized in Table~\ref{tab:1KNNValues}, where the central tendency and dispersion across the five samples are represented using median and range, respectively. 

\begin{table}
\begin{center}
\caption{1-KNN classification accuracy results (median $\pm$ range values across 5 samples) for the baseline and modified degradation models (50\% ideal benchmark).}
\label{tab:1KNNValues}
\begin{tabular}{lrr}
\toprule
Metric & Baseline & Modified \\
\midrule

Combined Classification Accuracy & $98.7\pm0.5\%$ & $63.0\pm7.1\%$ \\
Real Sample Accuracy   & $98.9\pm0.5\%$ & $54.8\pm11.7\%$ \\
Synthetic Sample Accuracy & $98.4\pm0.1\%$ & $71.3\pm6.4\%$ \\
\bottomrule
\end{tabular}
\end{center}
\end{table}

As noted in Table~\ref{tab:1KNNValues}, the modified model greatly improves the resulting 1-KNN accuracy towards the ideal performance metric of 50\%, with the combined (i.e.: real and synthetic sample) median classification accuracy reduced by 35.7\%. For the modified model, classification accuracy is considerably higher than real sample accuracy, indicating that the synthetic samples tend to be clustered in the eye tracking signal quality vector space. 

\section{Conclusions and Future Work}

A modified additive white Gaussian noise model for transforming the signal quality of previously captured eye tracking data was proposed and demonstrated herein. The model was shown to improve the resemblance of the synthetically-generated data to the target data repository on both a per-metric and aggregate basis using visual assessment of distributions and a 1-NN classification assessment, respectively. Namely, as shown in Figures \ref{fig:PrecFig} and \ref{fig:AccFig}, the percentile-matching technique introduced herein produces a distributional structure in the signal quality metrics with significantly enhanced similarity to the target data set when compared to the baseline transformation model. We hypothesize that the increased similarity between the synthetic and target data will better support data augmentation strategies for machine learning-based approaches within eye movement signal processing, including classification ~\cite{nuraini_machine_2021}, prediction~\cite{morales_saccade_2021}, and biometrics~\cite{lohr_metric_2020}. 

Moreover, the per-recording assessment shown herein further expands the literature by generalizing a technique that was previously applied to synthetically generated image data using generative adversarial networks to the signal domain. This technique is particularly valuable for the assessment of synthetic data generation techniques for machine learning applications, as it provides an application-agnostic framework to assess the efficacy of the proposed transform. Namely, by leveraging this approach, synthetic data generation algorithms can be quantitatively assessed without the additional burden of having to retrain the learned model for the augmented data set, thereby accelerating the iterative design cycle. While application-based metrics still serve as the ultimate assessment tool for a given target application, we hypothesize that the algorithm proposed herein can provide value to synthetic data designers by offering an alternative tool for rapid assessment during the preliminary design stage.

Subsequent investigations of the proposed model should focus on improving the resemblance between the spatial accuracy of the synthetic and real target data sets. This could be achieved by modifying the algorithm to allow for some consideration of the screen location of the fixation within the workflow, along with using a magnitude selection scheme which more accurately reflects the empirical distribution of this metric within the target data set. Modification of this portion of the model would better reflect the complexity of the distributions of spatial accuracy observed in real data (e.g. ~\cite{friedman2021angular}). 

In addition, future work should explore enhancing the richness of the proposed assessment workflow through the introduction of additional features within the summary feature vector. For example, including features which stereotype various eye movement dynamics (i.e.: adherence to various main sequence relationships for saccades, etc.) would help improve the validity of the assessment. Exploration of the relationship between the application-agnostic assessment mechanisms proposed herein and application-level performance (e.g. improved eye movement classification accuracy/generalization resulting from using this model for data augmentation, etc.) should also be conducted. 

\section{Acknowledgments}
Authors Samantha Aziz and Dillon Lohr are recipients of National Science Foundation Graduate Research Fellowship under Grant No. DGE-1840989 and DGE-1144466, respectively. Any opinion, findings, and conclusions or recommendations expressed in this material are those of the authors(s) and do not necessarily reflect the views of the National Science Foundation.

\bibliographystyle{plain}
\bibliography{references}  

\begin{thebibliography}{10}

\bibitem{ashraf2018eye}
Hajra Ashraf, Mikael~H Sodergren, Nabeel Merali, George Mylonas, Harsimrat
  Singh, and Ara Darzi.
\newblock Eye-tracking technology in medical education: A systematic review.
\newblock {\em Medical teacher}, 40(1):62--69, 2018.

\bibitem{bazrafkan2015eye}
Shabab Bazrafkan, Anuradha Kar, and Claudia Costache.
\newblock Eye gaze for consumer electronics: Controlling and commanding
  intelligent systems.
\newblock {\em IEEE Consumer Electronics Magazine}, 4(4):65--71, 2015.

\bibitem{coey2012structure}
Charles~A Coey, Sebastian Wallot, Michael~J Richardson, and Guy Van~Orden.
\newblock On the structure of measurement noise in eye-tracking.
\newblock {\em Journal of Eye Movement Research}, 5(4), 2012.

\bibitem{friedman2021angular}
Lee Friedman, Dillon Lohr, Timothy Hanson, and Oleg~V Komogortsev.
\newblock Angular offset distributions during fixation are, more often than
  not, multimodal.
\newblock {\em Journal of Eye Movement Research}, 14(3), 2021.

\bibitem{gardony2020eye}
Aaron~L Gardony, Robert~W Lindeman, and Tad~T Bruny{\'e}.
\newblock Eye-tracking for human-centered mixed reality: promises and
  challenges.
\newblock In {\em Optical Architectures for Displays and Sensing in Augmented,
  Virtual, and Mixed Reality (AR, VR, MR)}, volume 11310, page 113100T.
  International Society for Optics and Photonics, 2020.

\bibitem{griffith2020gazebase}
Henry Griffith, Dillon Lohr, Evgeny Abdulin, and Oleg Komogortsev.
\newblock Gazebase: A large-scale, multi-stimulus, longitudinal eye movement
  dataset.
\newblock {\em arXiv preprint arXiv:2009.06171}, 2020.

\bibitem{griffith_gazebase_figshare2020}
Henry Griffith, Dillon Lohr, and Oleg~V. Komogortsev.
\newblock {GazeBase} data repository, figshare, 2021.

\bibitem{holmqvist2012eye}
Kenneth Holmqvist, Marcus Nystr{\"o}m, and Fiona Mulvey.
\newblock Eye tracker data quality: what it is and how to measure it.
\newblock In {\em Proceedings of the symposium on eye tracking research and
  applications}, pages 45--52, 2012.

\bibitem{imaoka2020assessing}
Yu~Imaoka, Andri Flury, and Eling~D de~Bruin.
\newblock Assessing saccadic eye movements with head-mounted display virtual
  reality technology.
\newblock {\em Frontiers in Psychiatry}, 11:922, 2020.

\bibitem{leigh2015neurology}
R~John Leigh and David~S Zee.
\newblock {\em The neurology of eye movements}.
\newblock Contemporary Neurology, 2015.

\bibitem{Lohr2022}
Dillon Lohr, Samantha Aziz, Lee Friedman, and Oleg~V Komogortsev.
\newblock Gazebasevr, a large-scale, longitudinal, binocular eye-tracking
  dataset collected in virtual reality, 2022.

\bibitem{lohr_metric_2020}
Dillon Lohr, Henry Griffith, Samantha Aziz, and Oleg Komogortsev.
\newblock A metric learning approach to eye movement biometrics.
\newblock In {\em 2020 {IEEE} {International} {Joint} {Conference} on
  {Biometrics} ({IJCB})}, pages 1--7. IEEE, 2020.

\bibitem{lohr2019evaluating}
Dillon~J Lohr, Lee Friedman, and Oleg~V Komogortsev.
\newblock Evaluating the data quality of eye tracking signals from a virtual
  reality system: Case study using smi's eye-tracking htc vive.
\newblock {\em arXiv preprint arXiv:1912.02083}, 2019.

\bibitem{morales_saccade_2021}
Aythami Morales, Francisco~M. Costela, and Russell~L. Woods.
\newblock Saccade {Landing} {Point} {Prediction} {Based} on {Fine}-{Grained}
  {Learning} {Method}.
\newblock {\em IEEE Access}, 9:52474--52484, 2021.

\bibitem{nuraini_machine_2021}
Annis Nuraini, Suatmi Murnani, Igi Ardiyanto, and Sunu Wibirama.
\newblock Machine {Learning} in {Gaze}-{Based} {Interaction}: {A} {Survey} of
  {Eye} {Movements} {Events} {Detection}.
\newblock In {\em 2021 {International} {Conference} on {Computer} {System},
  {Information} {Technology}, and {Electrical} {Engineering} ({COSITE})}, pages
  150--155. IEEE, 2021.

\bibitem{otero2014unsupervised}
Jorge Otero-Millan, Jose L~Alba Castro, Stephen~L Macknik, and Susana
  Martinez-Conde.
\newblock Unsupervised clustering method to detect microsaccades.
\newblock {\em Journal of vision}, 14(2):18--18, 2014.

\bibitem{raynowska2018validity}
Jenelle Raynowska, John-Ross Rizzo, Janet~C Rucker, Weiwei Dai, Joel
  Birkemeier, Julian Hershowitz, Ivan Selesnick, Laura~J Balcer, Steven~L
  Galetta, and Todd Hudson.
\newblock Validity of low-resolution eye-tracking to assess eye movements
  during a rapid number naming task: performance of the eyetribe eye tracker.
\newblock {\em Brain injury}, 32(2):200--208, 2018.

\bibitem{sr2010eyelink}
SR~Research.
\newblock Eyelink 1000 user’s manual, version 1.5. 2, 2010.

\bibitem{snegireva2018eye}
N~Snegireva, W~Derman, J~Patricios, and KE~Welman.
\newblock Eye tracking technology in sports-related concussion: a systematic
  review and meta-analysis.
\newblock {\em Physiological measurement}, 39(12):12TR01, 2018.

\bibitem{st1973signal}
Gaetan~J St-Cyr.
\newblock Signal and noise in the human oculomotor system.
\newblock {\em Vision research}, 13(10):1979--1991, 1973.

\bibitem{tukey1977exploratory}
John~W Tukey et~al.
\newblock {\em Exploratory data analysis}, volume~2.
\newblock Reading, Mass., 1977.

\bibitem{wedel2017review}
Michel Wedel and Rik Pieters.
\newblock A review of eye-tracking research in marketing.
\newblock {\em Review of marketing research}, pages 123--147, 2017.

\bibitem{xu2018empirical}
Qiantong Xu, Gao Huang, Yang Yuan, Chuan Guo, Yu~Sun, Felix Wu, and Kilian
  Weinberger.
\newblock An empirical study on evaluation metrics of generative adversarial
  networks.
\newblock {\em arXiv preprint arXiv:1806.07755}, 2018.

\bibitem{zemblys2018using}
Raimondas Zemblys, Diederick~C Niehorster, Oleg Komogortsev, and Kenneth
  Holmqvist.
\newblock Using machine learning to detect events in eye-tracking data.
\newblock {\em Behavior research methods}, 50(1):160--181, 2018.

\end{thebibliography}






\end{document}